\documentclass[aps,prl,floatfix,twocolumn,showpacs]{revtex4} 
\usepackage{graphicx} \usepackage{amsmath} \usepackage{amssymb} 
\usepackage{dcolumn}
\usepackage[sort&compress]{natbib}
\newcommand{\BEQ}{\begin{equation}}
\newcommand{\EEQ}{\end{equation}}
\newcommand{\BEA}{\begin{eqnarray}}
\newcommand{\EEA}{\end{eqnarray}}

\renewcommand{\d}{{\rm d}}

\newcommand{\si}{\hat{\sigma}}
\newcommand{\siz}{\hat{\sigma}_z}
\newcommand{\six}{\hat{\sigma}_x}
\newcommand{\siy}{\hat{\sigma}_y}

\newcommand{\sipm}{\hat{\sigma}_{\pm}}

\newcommand{\om}{\omega}

\newcommand{\e}{\hat{\eta }}
\newcommand{\ha}{\hat{a}}

\newcommand{\X}{\hat{X}}

\newcommand{\taut}{\delta}

\newcommand{\half}{\frac{1}{2}}

\newcommand{\CE}{{\cal E}} 
\newcommand{\CP}{{\cal P}}

\newcommand{\HH}{\hat{H}}

\newcommand{\tr}{{\rm tr}}
\newcommand{\G}{F}
\begin{document} 
\draft
\title
{Bath Assisted Cooling of Spins}
\date{\today}
\author{A.E. Allahverdyan$^{1,2)}$,
R. Serral Graci\`a$^{1)}$
and Th.M. Nieuwenhuizen$^{1)}$}
\address{
$^{1)}$ Institute for Theoretical Physics,
University of Amsterdam,
Valckenierstraat 65, 1018 XE Amsterdam, The Netherlands,\\
$^{2)}$Yerevan Physics Institute,
Alikhanian Brothers St. 2, Yerevan 375036, Armenia
}

\begin{abstract}
A suitable sequence of sharp pulses applied to a spin coupled to a
bosonic bath can cool its state, i.e.,
increase its polarization or ground state
occupation probability. Starting
from an unpolarized state of the spin in equilibrium with the bath, one
can reach very low temperatures or sizeable polarizations within a time
shorter than the decoherence time.  Both the bath and external fields
are necessary for the effect which comes from the backreaction of the
spin on the bath.  This method can be applied to cool at once a
disordered ensemble of spins. 
Since the bath is crucial for this mechanism, the cooling limits
are set by the strength of its interaction with the spin(s).

\end{abstract}

\maketitle

Cooling, i.e. obtaining relatively pure states from mixed ones, is of
central importance in fields dealing with quantum features of
matter. Laser cooling of motional states of atoms is nowadays a known
achievement \cite{lasercooling}.  The related problem of cooling spins
is equally known: it originated as an attempt to improve the
sensitivity of NMR/ESR spectroscopy
\cite{abo,ole,slichter,exp,algol,turk}, since in experiments the
signal strength is proportional to the polarization.  Recently it got
renewed attention due to realizations of setups for quantum computers
\cite{chuang}.  The very problem arises since the most direct methods
of cooling spins, such as lowering the temperature of the whole sample
or applying strong dc fields, are not feasible or not desirable,
e.g. in biological applications of NMR.  Indeed, at temperature $T=1$K
and magnetic field $B=1$T the equilibrium polarization of a proton is
only $\tanh\frac{\hbar\mu B}{2k_{\rm B}T} = 10^{-3}$ since the ratio
$\mu=\frac{\rm frequency}{\rm field}$ is equal to $42$ MHz/T. For an
electron $\mu$ is $10^3$ times larger and for $^{15}$N it is $10$
times smaller.  The weak polarization can be often compensated by a
large number of spins, but for some NMR-isotopes the natural abundance
is too low ($0.36\%$ for $^{15}$N).

Over the years, several methods were proposed to attack the problem of
small polarizations. The polarization is generally increased via a
dynamical process and it is used before relaxing back to equilibrium
\cite{abo,ole,exp,slichter,turk,algol}.  Specially known are methods
where a relatively high polarization is transferred from one place to
another, e.g. from electronic to nuclear spins
\cite{abo,ole,exp,slichter,turk}. In this respect electronic spins play the
same role as the zero-temperature bath of vacuum modes employed for
laser cooling of atoms \cite{lasercooling} (this bath is typically
inadequate for cooling nuclear spins, but can be employed to study
cooling of atomic few-level systems in the context of optimal control
theory \cite{tannor}).  Polarization transfer was studied in
various settings both theoretically and experimentally
\cite{abo,ole,slichter,exp,turk}. However, this scheme is limited
|besides requiring an already existing high polarization| by the
availability and efficiency of the transfer interaction.  A related
method, polarization compression, consists in manipulating a set of
$n$ spins in such a way that the
polarization of one spin is increased at the expense of decreasing the
polarization of the remaining $n-1$ spins. These spoiled ones can be
recycled and used again \cite{algol}. 
Since spins are cooled one by one, a long time 
and carefully designed inter-spin interactions are needed
for cooling a large ensemble.

Here we propose a mechanism of cooling which only uses the most
standard setting of NMR or ESR physics \cite{abo,ole,slichter}:
spins-$\frac{1}{2}$ under the action of external field pulses coupled
to a thermal bath at the same temperature. The bath is needed because
external fields alone cannot achieve cooling \cite{nocooling}.
However, we assume neither that the bath is under any direct control,
nor special constraints on the bath-spin interaction: it is the
standard one, widely studied in the context of decoherence.  We show
that, rather than being a hindrance in quantum system manipulations,
the bath is capable of producing ordered effects on the
spin, which can cool it down to very low temperatures ($\sim 1\mu$K
for a proton) in a finite time.  Two factors are crucial: the
backreaction of the spin on the bath and the generation of transversal
components (coherences) during the cooling process.  Since the effect
is generated via the bath, one can cool {\it at once} a completely
disordered ensemble of spins.

The model we study is well known \cite{ms,lu,viola,lidar,ANjpa}: a
spin-$\half$ with
energy levels $\pm\half{\hbar\Omega}$ couples to a bath,
modeled by a set of harmonic oscillators with 
creation and annihilation operators $\ha_k^{\dagger}$ and $\ha_k$. 
The total Hamiltonian reads
\BEA
\label{ham}
\HH=\frac{\hbar\Omega}{2}\si _z+
{\sum}_k\hbar\om _k\ha^{\dagger}_k\ha _k+\frac{\hbar\si_z}{2}\X,
\quad [\ha _l,\ha ^{\dagger}_k]=\delta _{kl}.
\EEA
Here $\om_k$ are the bath frequencies, $\si _{x,y,z}$ the Pauli
operators, and $\X={\sum} _kg_k(\ha _k^{\dagger}+\ha _k)$ is the
collective coordinate of the bath.  The interaction is chosen
assuming that the ${\cal T}_1$-time, connected to relaxation of the
average $\langle\siz\rangle$ is very large (infinite) \cite{abo,ole,slichter}.
The $g_k$ are couplings parametrized via the spectral density
function $J(\om)$:
\BEA
J(\om)={\sum}_kg^2_k\, \delta(\om-\om_k).
\label{ja1}
\EEA
In the thermodynamic limit the bath modes are dense 
and $J(\om)$ becomes a smooth function
determined by the physics of the system-bath
interaction \cite{ms}.
The oscillators can represent real
phonons or stand for an effective description of a rather general
class of thermal baths \cite{ms}. 

Let us recall how the model (\ref{ham}) 
is solved \cite{lu}: $\siz$ is conserved, while
$\ha _k(t)=
e^{-i\om _kt}\ha _k(0)+\frac{g_k\si _z}{2\om _k}\,\left(
e^{-i\om _kt}-1
\right)$.
This leads along with Eqs.~(\ref{ham}, \ref{ja1}) to
\BEA
\label{kant}
&&\X(t)=\e (t)-\si_z\dot{\G}(t),
\\
\label{eta=}\label{engels}
&&\e  (t)\equiv
{\sum}_kg_k[\ha _k^{\dagger}(0)e^{i\om _kt}+\ha _k(0)e^{-i\om _kt}],
\\ \label{marx}
&&\G(t)\equiv
\int_0^\infty \frac{\d \om}{\om}J(\om)(t-\frac{\sin \om t}{\om}),
\EEA
where $\e(t)$ is the quantum noise operator, and where
$\dot{\G}(t)\equiv \frac{\d}{\d t}\G(t)$
quantifies the {\it backreaction} of the spin on the 
collective operator of the bath. This effect, not relevant for decoherence as such,
is crucial for our purposes.

We assume that at the initial time $t=0$ the common density matrix of
the bath and the spin is \emph{factorized}:
\BEA
\label{fedor}
\rho (0)=
\frac{e^{-\beta \HH_0}}{\tr\,e^{-\beta \HH_0}},\quad
\HH_0=\frac{1}{2}{\hbar\Omega}\si _z+
{{\sum}}_k\hbar\om _k\ha^{\dagger}_k\ha _k,
\EEA
where $T\equiv 1/\beta$ is the common temperature ($k_{\rm B}=1$).  
$\rho(0)$ describes the spin prepared
independently from the bath and then brought in contact
with it at $t=0$, e.g., by injection of the spin
into a quantum dot or by creation of an exciton through external radiation.  

As follows from Eq.~(\ref{fedor}),
$\e(t)$ is a Gaussian operator with $\langle \e (t)\rangle=0$ 
and time-ordered correlator
$\langle \overleftarrow{\e  (t) \e  (0)}\rangle
=\ddot\xi(t)-i\ddot \G (t),~~t>0$,
where $\langle...\rangle$ is taken over the initial state (\ref{fedor}),
and where
\BEA
\label{during}
\xi(t)\equiv \int _0^{\infty}\d \om\,J(\om)
\frac{1-\cos\omega t}{\om^2}\,\coth\frac{\hbar\om}{2T}.
\EEA
The Heisenberg equation of the spin,
$\hbar\dot{\si}_\pm=i[\HH,\si_\pm]$ with
$\sipm =\si _x\pm i\,\si _y$, 
$\si _z\sipm =\pm\sipm$, is solved as 
\BEA
\label{k4}
\hat{\sigma}_\pm (t)=e^{\pm\, i\Omega (t-t_0)-i\G(t-t_0)
}\,\overleftarrow{e}^{\,\pm\,
i\int_{t_0}^{t}\,\d s\,\e (s)
}\,\hat{\sigma}_\pm (t_0),
\EEA
where $\overleftarrow{e}$ is the time-ordered exponent.
Defining 
$\CE_t\hat{A}\equiv e^{{it}\HH/{\hbar}}\hat{A}e^{-{it}\HH/{\hbar}}$
one derives
\BEA
\label{suki}
\CE_t\overleftarrow{e}^{\,\pm\,
i\int_{t_1}^{t_2}\,\d s\,\e (s)
}=\overleftarrow{e}^{\,\pm\,
i\int_{t_1+t}^{t_2+t}\,\d s\,\e (s)
}e^{\pm i\siz \chi(t_1,t_2,t)},~~\\
\chi(t_1,t_2,t)\equiv
F(t_2)-F(t_1)+F(t_1+t)-F(t_2+t),
\label{muki}
\\
\langle
\overleftarrow{e}^{\,\pm\,
i\int_{t_1}^{t_2}\,\d s\,\e (s)
}\rangle=
e^{-\xi(t_2-t_1)+iF(t_2-t_1)}.~~~
\label{ormo}
\EEA
Eq.~(\ref{ormo}) is the standard formula for 
the average of a Gaussian operator. 
The factor $e^{-\xi(t)}$ leads to
decoherence \cite{lu}, since due to
Eqs.~(\ref{k4}, \ref{ormo}),
$\langle\hat{\sigma}_\pm (t)\rangle=e^{-\xi(t)\pm i\Omega t}
\langle\hat{\sigma}_\pm (0)\rangle$ 
for a general factorized initial state. 
In this simplest situation the backreaction factor $F$,
properly obtained already in \cite{lu,viola},
cancels out. In general, $F$ can shift the spin's frequency
$\Omega$ as seen below.

The action of external fields on the spin amounts
to a time-dependent Hamiltonian $\HH(t)=\HH+
\vec{h}(t)\vec{\si}$.
In the {\it pulsed} regime \cite{abo,ole,slichter,viola,lidar,ANjpa}
$\vec{h}(t)$ differs from zero only for very
short intervals of time $\delta$ being there very large,
$\vec{h}(t)\delta\sim 1$, to achieve a finite effect.
As a consequence, terms $\propto\siz$ in $\HH$ can be neglected
during the time-interval $\taut$.
A single pulse can perform an arbitrary unitary transformation
in the space of the spin (rotation of the 
Bloch vector $\langle \vec{\hat{\sigma}}\rangle$). 
We parametrize it as 
$\hat{U}\equiv e^{{i\,\delta}\,\vec{h}(t)\vec{\si}/\hbar}$,
($0\leq \phi,\psi\leq 2\pi$, $0\leq\theta\leq\pi/2$) 
\BEA
\label{para}
\hat{U}
=\left(\begin{array}{rr}
e^{-i\phi}\cos\vartheta & -e^{-i\psi}\sin\vartheta \\
e^{i\psi}\sin\vartheta& e^{i\phi}\cos\vartheta
\end{array}\right),\;
\CP\hat{A}\equiv \hat{U}\hat{A}\hat{U}^\dagger,
\EEA
As a first example, we take ohmic interaction \cite{ms}:
\BEA
\label{ja2}
J(\om)=\gamma\,\om\,e^{-\omega/\Gamma}.
\EEA
where $\gamma $ is a dimensionless coupling constant, 
and where $\Gamma$ (usually $\gg \Omega$) is the bath's response
frequency. Eqs.~(\ref{marx}, \ref{during}, \ref{ja2}) imply
$\xi(t)= \gamma \,
\ln \left [\frac{{\mathbf \Gamma} ^2\left (1+\Theta
\right )
\,\,\sqrt{1+\Gamma^2t^2}}
{{\mathbf \Gamma}\left (1+\Theta
-i\Theta \,\Gamma \,t\right )
{\mathbf \Gamma} \left (1+\Theta
+i\Theta\,\Gamma \,t\right )}
\right],$
\BEA
\G(t)=\gamma\left[
\Gamma t-\arctan(\Gamma t)\right],\quad
\Theta\equiv \frac{k_{\rm B}T}{\hbar\Gamma},~~
\EEA
where ${\mathbf \Gamma}$ is Euler's function and $\Theta$ is the
dimensionless temperature. 
The time-scale ${1}/{\gamma\Gamma}$ of
backreaction $\G(t)$ is {\it temperature-independent} as opposed
to the decoherence time.
For low temperatures
$\Theta\ll 1$: $e^{-\xi(t)}=(1+t^2\Gamma^2)^{-\gamma/2}$, while for
$\Theta\gtrsim 1$, $e^{-\xi(t)}$ starts as a gaussian, but continues as
$e^{-t/{\cal T}_2}$ with ${\cal T}_2 ={\hbar}/({2\gamma
T})$ \cite{lu}.

{\it Cooling} amounts to make the final polarization
$\langle\siz\rangle_{\rm f}$ more negative than the initial one, 
$\langle\siz\rangle_{\rm i}=-\tanh\frac{1}{2}\beta\hbar\Omega$ 
(Eq.~(\ref{fedor}) implies $\langle\six\rangle_{\rm
i}=\langle\siy\rangle_{\rm i}=0$).  A single pulse cannot achieve
cooling since it sees the initial local equilibrium state of the spin,
and then according to the no-cooling principle \cite{nocooling} it can
only heat the spin's state up: for an arbitrary pulse $\CP_1$ applied at
time $t$, $\langle\siz\rangle_{\rm f}\equiv \langle
\CE_t\,\CP_1\,\siz\rangle=\langle\siz\rangle_{\rm i} \cos 2\vartheta_1
\geq \langle\siz\rangle_{\rm i}$ (recall $\langle\siz\rangle_{\rm
i}\lesssim 0)$. Thus we have to employ at least two pulses. 
The final polarization after one pulse at
$t$ and one at $t+\tau$, $P=|\langle\siz\rangle_{\rm f}| =|\langle
\CE_t\CP_1\CE_\tau\,\CP_2\siz\rangle|$, 
reads from Eqs.~(\ref{k4}-\ref{para}):
\BEA
\label{koshik1}
\langle\siz\rangle_{\rm f}
=\langle\siz\rangle_{\rm i}
\cos 2\vartheta_1\cos 2\vartheta_2\,
+s_2\sin 2\vartheta_1\sin 2\vartheta_2,\\
s_2=-e^{-\xi(\tau)}\Re\left\{
e^{i\Omega \tau+i\alpha_2}\,\left(\,\langle\siz\rangle_{\rm i} 
\cos \chi + i \sin \chi \right) \right\},
\label{koshik2}
\EEA
where $\chi=\chi(0,t,\tau)$ was defined in Eq. (\ref{muki}), and
$\alpha_2=\psi_1-\psi_2-\phi_1-\phi_2$ arises
from Eq. (\ref{para}).  There
are now {\it two} factors that come from the bath: $e^{-\xi(\tau)}$ in
$s_2$ accounts for the decoherence in the time-interval $(t,t+\tau)$
of transversal terms generated by the first pulse, while $\chi$ is the
backreaction factor from Eqs.~(\ref{suki}, \ref{muki}).  

Though the finite-$t$ situation can be of its own
interest, for all results below we set $t\Gamma\gg 1$
(a mild condition, since $1/\Gamma$ is typically the shortest
time-scale),
since this makes the outcome independent on the details of the initial
state preparation.
In this {\it ergodic limit}, the initial condition 
$\rho(0)\propto e^{-\beta \HH_0}$ 
defined by Eq.~(\ref{fedor})
is equivalent to the overall
equilibrium preparation $\rho_{\rm eq}(0)\propto
e^{-\beta \HH}$ \cite{arn}.

\begin{figure}
\includegraphics[width=0.49\linewidth]{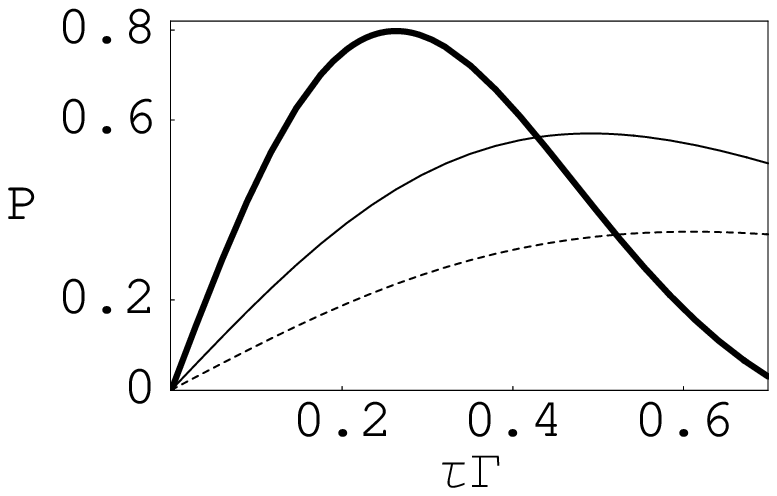}%
\hfill%
\includegraphics[width=0.49\linewidth]{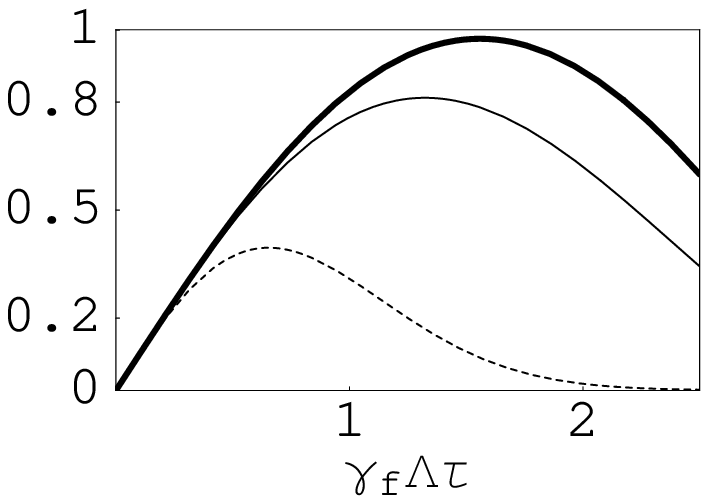} 
\caption{
Final polarization $P=|\langle\siz\rangle_{\rm f}|$ 
given by Eqs.~(\ref{koshik1}, \ref{koshik2})
versus dimensionless time between two optimal
pulses. 
Left: {\it ohmic interaction}.
Bold curve: $\gamma=5$, $\Theta=0.2$.
Normal: $\gamma=2$, $\Theta=0.5$.
Dashed: $\gamma=1$, $\Theta=1$.
Right: {\it $1/f$ interaction}.
$\gamma_f> 10^3$ and ${\Theta_f}/{\gamma_f}=0.01,\,0.1,\,1$
(bold, normal, dashed). 
In both figures 
$\langle\siz\rangle_{\rm i}$ and
$\Omega$ are neglegible.} 
\label{1} 
\label{2}
\end{figure}

In Eq.~(\ref{koshik2}), $s_2$
can always be made negative by tuning
$\alpha_2$. Minimizing $\langle\siz\rangle_{\rm f}$ over
$\vartheta_1,\vartheta_2$ produces ${\rm min}\left
[\langle\siz\rangle_{\rm i},s_2 \right]$. If the initial polarization
is already high, $|\langle\siz\rangle_{\rm i}|>|s_2|$, 
no pulses should be applied, since they only heat
the spin up. However, in the relevant situation 
$\langle\siz\rangle_{\rm i}\simeq 0$, the minimum
$\langle\siz\rangle_{\rm f}=s_2$ is reached for
$\vartheta_1=\vartheta_2 =\frac{\alpha_2}{2}=\frac{\pi}{4}$.  
Altogether, using Eq.~(\ref{muki}) 
and ${\Omega}\ll{\Gamma}$, yields 
\BEA
\langle\siz\rangle_{\rm f}=s_2=-
e^{-\xi(\tau)}
\sin\left[
\gamma\arctan(\tau\Gamma)
\right].
\label{koshik20}
\EEA
The choice of optimal pulses, which have to be coherence generating, 
can be a $\frac{\pi}{2}$-pulse along the $x$-axis followed by a
$-\frac{\pi}{2}$-pulse along the $y$-axis:
$\CP_1\hat{\sigma}_{z,x}=\hat{\sigma}_{y,x}$,
$\CP_1\hat{\sigma}_{y}=-\hat{\sigma}_{z}$, and
$\CP_2\hat{\sigma}_{z,y}=\hat{\sigma}_{x,y}$,
$\CP_2\hat{\sigma}_{x}=-\hat{\sigma}_{z}$.  Fig. \ref{1} and Table I
show that
${\rm max}_\tau|\langle\siz\rangle_{\rm f}|$ can approach its maximal
value $1$. Transversal components generated by the two pulses
will decay after a time ${\cal T}_2$, and 
the spin will be described by a Gibbsian at
a temperature lower than the initial $T$.

The origin of this cooling effect is in shifting the spin's frequency
$\Omega$ by the factors $F\sim\gamma\Gamma$ 
and $\chi$, recall Eqs.~(\ref{k4}, \ref{muki}),
which arise from the (via the pulses) enhanced 
backreaction of the spin on the collective
coordinate $\hat{X}$ of the bath. The generation of coherences by the first
$\frac{\pi}{2}$-pulse is necessary to couple $\si_z$ to the bath,
which so to say ``thermalizes'' 
$\langle\si_z\rangle$ under the shifted frequency.
Cooling is achieved via the proper pulses, still
it decreases with $\gamma$ (weaker backreaction) and with
$\beta$ (larger decoherence).
The time $\tau$ between two pulses can be neither too short
($\chi$ is visible on the time-scale $1/(\gamma\Gamma)$),
nor too long, since otherwise decoherence will diminish
the influence of the first pulse.
\begin{table}
\caption{
Maximum value of $|\langle\siz\rangle|$ for $2$, $3$ and
$50\times 2$ pulses and various values of 
$\gamma$ and $\Theta=k_{\rm B}T/(\hbar\Gamma)$
(for the ohmic case).
The lowest two lines represent the final collective polarization 
$|m_{\rm f}|$ of a strongly disordered ensemble with $\Omega_0=0$
and arbitrary large disorder dispersion $d$
under spin-echo setup of 2 pulses plus a $\pi$ pulse 
and a sequence of $50$ such sets of pulses.
}
\begin{tabular}{|c||c|c||c|c||c|c|}
\hline 
$\gamma$ & 0.1 & 0.1 & 1 & 1. & 10. & 10. \\
\hline $\Theta$ &0.1 & 1. &0.1 & 1  & 0.1 & 1. \\
\hline
\hline $2$  & 0.1126 & 0.0826 & 0.4930 & 0.3516 & 0.8910 & 0.7931 \\
\hline $3$  & 0.2036 & 0.1369 & 0.7250 & 0.4732 & 0.9550 & 0.9025 \\
\hline  $50\times 2$  &0.2779 & 0.1895 &0.6948 & 0.5430  & 0.9509 & 0.8995 \\
\hline
\hline $2+\pi$ & 0.0834& 0.0604 & 0.2498 & 0.2026  & 0.4634  & 0.4155 \\
\hline $50\times [2+\pi]$ & 0.1396 & 0.1055 & 0.3119 & 0.2639 & 0.5214
& 0.4834  \\
\hline
\end{tabular}
\label{tab1}
\end{table}
As shown in Table I, the cooling improves by {\it i)}
applying three successive pulses and optimizing over their parameters;
{\it ii)} applying two pulses, waiting for a time $\geq {\cal T}_2$,
so that the transversal components decay, $\langle\sipm\rangle\to 0$,
applying another two pulses, and so on $n$ times.  The final $|\langle\siz
\rangle|$ is maximized over all free parameters.
It appears that this numerical maximization can be done locally,
i.e., by maximizing the output $|\langle\siz\rangle|$ after each pair
of pulses. This ``greedy'' optimization shows up also in cooling
via a zero-temperature bath \cite{tannor}.

{\it Inhomogeneous broadening.}  Many experiments in NMR/ESR are not done
with a single spin, but with an ensemble of non-interacting spins
having random frequencies $\Omega$ due to action of their environment
or due to inhomogeneous external field
\cite{abo,slichter}. 
The collective variables are obtained by averaging 
the corresponding expressions for a single spin: $m_{\rm
f}=\int\d\Omega\,P(\Omega)\langle\siz\rangle_{\rm f}$.  Assume that
the distribution of $\Omega$ is gaussian with average $\Omega_0$ and
dispersion $d$: $ P(\Omega)\propto e^{-(\Omega-\Omega_0)^2/(2d)}$.
Averaging over $P(\Omega)$ the term $e^{i\Omega\tau}$ in
Eq.~(\ref{koshik2}) produces a factor $\sim e^{-d\tau^2/2}$, a strong
decay on times ${\cal T}_2^*\propto 1/\sqrt{d}$. After this decay,
$s_2\to 0$ and any two pulses will only heat the ensemble up as seen
from Eq.~(\ref{koshik1}).

It is however possible to employ the spin-echo phenomenon and cool,
i.e., increase the {\it collective}
final polarization $|m_{\rm f}|$ as compared to
the initial $|m_{\rm i}|$, even for a {\it completely} disordered ensemble with
${\cal T}_2^*$ being very short: Apply 
{\it precisely in the middle of the two pulses} an additional $\pi$-pulse in
$x$-direction: $\CP_\pi\si_{y,z}=-\si_{y,z}$,
$\CP_\pi\si_{x}=\si_{x}$, and work out
$m_{\rm f}=
\int\d\Omega\,P(\Omega)\langle 
\CE_t\CP_1\CE_\tau\CP_\pi \CE_\tau\,\CP_2\siz\rangle$:
\begin{gather}
\label{koshik3}
m_{\rm f}
=-m_{\rm i}\cos 2\vartheta_1\cos 2\vartheta_2
+s_3\sin 2\vartheta_1\sin 2\vartheta_2,\\
s_3=-e^{-4\xi(\tau)+\xi(2\tau)}
\Re\left\{
e^{i\alpha_3}\left( m_{\rm i}\cos\chi_3-i\sin\chi_3
\right)\right\},
\label{koshik4}\\
\chi_3=\chi(0,\tau,t)-\chi(\tau,2\tau,t),
\label{koshik5}
\end{gather}
where $\alpha_3=\phi_1-\psi_1-\phi_2-\psi_2$.  As compared to
(\ref{koshik2}), both the decoherence $e^{-4\xi(\tau)+\xi(2\tau)}$ and
the backreaction $\chi_3$ term are different. In the
gaussian regime  $\xi\propto t^2$ decoherence is absent, while
the exponential regime  $\xi\propto t$ is unchanged, 
$e^{-4\xi(\tau)+\xi(2\tau)}=e^{-\xi(2\tau)}$ \cite{viola,lidar,japan1/f}.  
Due to the $\CP_\pi$
pulse, the ${\cal T}_2^*$-decay has been eliminated, no term like 
$e^{-i\Omega t}$ in Eq. (\ref{koshik2}) appears here.
Now $\Omega_0$ and $d$ enter only via $m_i$.
The structure of Eqs.~(\ref{koshik3}-\ref{koshik5})
is close to the one of Eqs.~(\ref{koshik1},\ref{koshik2}), and
the optimization over $\vartheta_1,\vartheta_2$ goes in the
same way. To facilitate comparison, we take $\Omega_0=0$,
thus $m_{\rm i}=0$, and disorder strength
$d$ {\it arbitrary} large. recalling in addition
Eqs.~(\ref{muki}, \ref{ja2}) and
the ergodic condition $t\Gamma\gg 1$, we get
\BEA
|m_{\rm f}|=e^{-4\xi(\tau)+\xi(2\tau)}
\sin\{\gamma\left[
2\arctan(\tau\Gamma)-\arctan(2\tau\Gamma)
\right]\,\},\nonumber
\label{koshik30}
\EEA
where we already inserted the optimal values
$\vartheta_1=\vartheta_2=-\frac{\alpha_3}{2}
=\frac{\pi}{4}$:
The choice of optimal pulses can be the same as for the two-pulse
scenario. The maximal $|m_{\rm f}|$ can exceed
$0.4$ for sufficiently strong coupling and/or low temperatures. 
The results improve by
applying a sequence of three (spin-echo) pulses separated 
from each other by a time
much larger than ${\cal T}_2$, see Table I .

A {\it $1/f$ spectrum}
is another relevant situation of the bath-spin interaction
recently observed in a two-level
system (spin) of charge states in Josephson-junction circuit (Cooper-pair
box) \cite{japan1/f}. 
The spin's interaction with the bath of background charges 
is modelled via \cite{japan1/f,lidar}
$J_f(\om)=\frac{b_{f}}{\om}\,e^{-\om/\Gamma}\,\theta (\om-\Lambda)$,
where $b_{f}$ is the coupling constant, $\theta$ is the step function,
and where $\Gamma$ and $\Lambda$ are, respectively, the largest and
smallest frequencies of bath's response.  
The upper frequency $\Gamma$ is not relevant: the
integrals in Eqs.~(\ref{marx}, \ref{during})  converge for
$\Gamma\to \infty$ \cite{lidar}.  Within the realization of \cite{japan1/f}:
$\Gamma\to\infty\,(\gg 10^{11}$Hz), $\Lambda=314$Hz, $\hbar\Omega=34\mu$eV,
while the relevant dimensionless coupling constant is
{\it very large} $\gamma_f\equiv{b_f}/{\Lambda^2}=2.3\times 10^{11}$.
Thus the cooling effect will be especially visible here.
The spin-boson model does not describe all the aspects
of the decoherence in Cooper-pair box; see \cite{ss} for more elaborated
approaches. Still some important features are reproduced adequately
\cite{japan1/f,lidar}, and this motivates us to
work out Eqs.~(\ref{marx}, \ref{during}) and apply
them to the two-pulse cooling situation described by Eqs.~(\ref{koshik1},
\ref{koshik2}) under the following conditions. {\it i)} 
high temperatures $\Theta_f={k_{\rm B}T}/{(\hbar\Lambda)}\gg 1$
(pessimistic case!);  {\it ii)} $\tau\Lambda\ll 1$ and $\gamma_f\gg 1$ 
(experimentally relevant regimes \cite{japan1/f});  {\it iii)}
$\langle\siz\rangle_{\rm i}\simeq 0$ due to large temperatures,
 {\it iv)} ${\Omega\tau}\ll{\gamma_f\Lambda}$ due to large 
$\gamma_f$. In analogy to Eq.~(\ref{koshik20}), 
the result is
($\alpha_2=\frac{\pi}{2}$): $
\langle\siz\rangle_{\rm f}=
e^{-\Theta_fy^2/\gamma_f }\sin\left(
y-\frac{\pi y^2}{4\gamma_f}
\right)$, where $ y\equiv\gamma_f\Lambda \tau$ is
a dimensionless time, and where we omitted terms
$O[{y^3}/{\gamma^2_f}]$.  As seen in 
Fig. \ref{1}, the polarization can increase from its original value
zero to $0.997$ for ${\Theta_f}/{\gamma_f}=0.01$, which in numbers of
Ref.~\cite{japan1/f} corresponds to $T=15$K.  

{\it In conclusion}, we described a new method for cooling spins due
to {\it common} action of external fields and a bosonic bath, starting
from the overall equilibrium state. The fields alone cannot cool
\cite{nocooling}, while the bath alone can generate only the standard
decoherence \cite{lu}. As compared to existing methods
\cite{abo,ole,slichter,exp,turk,algol,tannor}, the present one
assumes neither already existing high polarization
\cite{abo,ole,slichter,exp,turk}, nor controlled spin-spin or
bath-spin interactions \cite{algol}, nor a low-temperature bath
\cite{tannor}.  It works even for very weak dc fields and applies to
an ensemble of spins having completely random frequencies (strong
inhomogeneous broadening). We are not aware of other methods achieving
such a goal. The spins are cooled at once (not one by one) and the
cooling process takes a time shorter than ${\cal T}_2$.  Together with
the overall efficiency of the method, see the figures and the Table,
these features are encouraging for applications, e.g., in NMR
spectroscopy.  Our basic assumptions are a decoherence time ${\cal
T}_2$ much smaller than the energy relaxation time ${\cal T}_1$, and
the availability of sharp and strong pulses acting on the spin.  A
long ${\cal T}_1$ time characterizes other methods \cite{abo,ole},
while strong and short pulses were used for a clean demonstration of
the effect, which probably survives for other types of pulses.

The origin of the present mechanism lies in shifting the spin's
frequency due to backreaction of the spin on the bath. This dynamical
effect requires a non-perturbative
treatment of the bath-spin interaction and is usually missed by
markovian approaches \cite{lindblad}.
It operates on a specific time-scale and allows to cool the spin
provided the proper, coherence generating, sequence of external pulses is chosen. 
The cooling is efficient already for small-to-moderate bath-spin
couplings, and is especially visible for situations where a strong
bath-spin coupling is inherent ($1/f$-noise). In this
experimentally realized situation \cite{japan1/f}
the cooling mechanism is expected to be feasible.


We thank P.A. Bushev for interesting discussions on cooling.
This work was supported by FOM/NWO.

\end{document}